\documentclass[11pt]{article}
\title{\protect\LARGE\bf  
Extended Ensemble Monte Carlo}

\author{Yukito Iba \\ The Institute of Statistical Mathematics \\
iba@ism.ac.jp}

\date{}

\newcommand{\boldx}{\mbox{\boldmath $x$}}

\newcommand{\boldlam}{\mbox{\boldmath $\lambda$}}

\begin{document}

\maketitle
\begin{abstract}
``Extended Ensemble Monte Carlo''
is a generic term that indicates a set of
algorithms which are now popular in 
a variety of fields in  physics and statistical information processing.
Exchange Monte Carlo (Metropolis-Coupled Chain, Parallel Tempering),
Simulated Tempering (Expanded Ensemble Monte Carlo), and
Multicanonical Monte Carlo (Adaptive Umbrella Sampling)
are typical members of this family.
Here we give a cross-disciplinary survey of 
these algorithms with special emphasis on the great flexibility 
of the underlying idea. 
In Sec.~\ref{SS_art}, we discuss the background of
Extended Ensemble Monte Carlo. In
Sec.~\ref{SS_exg}, \ref{SS_temp} and \ref{SS_multi}, three types
of the algorithms, i.e., {Exchange Monte
Carlo}, {Simulated Tempering}, 
{Multicanonical Monte Carlo},
are introduced. 
In Sec.~\ref{SSS_rep}, we give
an introduction to Replica Monte Carlo algorithm by Swendsen
and Wang.
Strategies for the construction of 
special-purpose extended ensembles are discussed in 
Sec.~\ref{SS_new}. We stress that an extension is
not necessary restricted to the space
of energy or temperature. Even unphysical (unrealizable) 
configurations can be included in the ensemble, 
if the resultant fast mixing of the Markov chain offsets 
the increasing cost of the sampling procedure.
Multivariate
(multi-component) extensions are also useful in many examples.
In Sec.~\ref{SSS_msse}, we give a survey on
extended ensembles with a state space whose dimensionality 
is dynamically varying.  In the appendix, 
we discuss advantages and disadvantages of 
three types of extended ensemble algorithms. 
\end{abstract}

\vspace{0.5cm}
\begin{quote}
{\small \bf Keywords:} \\
{\sf
Extended Ensemble, Exchange Monte Carlo,
Simulated Tempering, Multicanonical Monte Carlo, Replica Monte Carlo,
Complexity Ladder, Bridge, Multivariate Extension
}
\end{quote}

\newpage

\sloppy
\tableofcontents

\newpage

\section{Introduction}

In this paper, we will give 
a survey on {\bf Extended Ensemble Monte Carlo} 
algorithms~\footnote{We choose `` Extended Ensemble Monte Carlo''
as a generic term to represent a family of
algorithms which we want to discuss here, {\it e.g.}, Exchange Monte 
Carlo, Multicanonical Monte Carlo, {\it etc}. Another term,
{\bf Generalized Ensemble Monte Carlo}, is used by some authors.
The term {\bf Expanded Ensemble Monte Carlo}, which we use here
in a more restricted meaning, can also be used
in the generic meaning. However, the original 
definition~\cite{expand}
 of Expanded Ensemble 
Monte Carlo seems not to cover Exchange Monte Carlo.}, which are
useful tools in computational physics and 
in the fields of statistical information
processing. Well-known algorithms in this family are
{\bf Exchange Monte
Carlo} ({Metropolis-Coupled Chain}, 
{Parallel Tempering})~\cite{KT90,IHM92,G91,GT95,I93,H96,Tesi},
 {\bf Simulated Tempering}
({Expanded Ensemble Monte Carlo})~\cite{expand,MP92} 
and {\bf Multicanonical Monte Carlo}
 ({Adaptive Umbrella Sampling})~\cite{M87,HVK92,BergN,BergC,Berg00}.
These approaches are characterized by {\it modification
of ensembles} sampled by the algorithm. 
In this respects, they contrast with other
attempts to overcome the limitation of conventional 
Dynamical Monte Carlo, i.e., {\it improved dynamics}
that preserve original ensembles~\cite{CMC,Kron} and
{\it improved algorithms} that maintain 
original dynamics~\cite{Novo}. 

These algorithms are useful for the studies of 
stochastic models in various fields of physics,
{\it e.g.}, spin models (Potts models~\cite{BergN,KW93}, spin glass
models~\cite{BergC,BC92b,H96,H99,HK99,NPR,MMZ01,PCA01,KR94,MOE,msse,SW86,SW88,H01}, 
random field models~\cite{MP92}, quantum spin models~\cite{qs}),
polymer models (lattice polymers~\cite{Tesi,expand96},
diblock copolymer~\cite{WYNP00}, lattice 
heteropolymers/proteins~\cite{UT,iba98,Chiken99,CK00,HO2}, 
off-lattice polymers~\cite{expand96,IrbackMD,EP95,EP96},
realistic protein/polypeptide models~\cite{HO,HO2,
Mitsutake01,SO01,NN97,NN98,ONHN99,HGON01,BK98}),
models of molecules in vacuum or 
water~\cite{M87,HVK92,BK97},
hard core fluid (solid)~\cite{expand,SB96,BWA,Mase99,Wilding01},
Lennard-Jones fluid~\cite{expand,Wnum,WildingR}, 
models of aqueous solution~\cite{LFL98,expandhttp},
Lennard-Jones clusters~\cite{CNFD00}, 
lattice gauge models~\cite{lathybrid,latexchange}, 
models of quantum gravity~\cite{qg}. 
They are also successfully used in 
statistical inference~\cite{GT95,LS98,ABH01} 
and combinatorics~\cite{PW98}.
Our aim here is, however, not to give a
list of references on this subject. Instead, we want to
discuss basic ideas behind algorithms and show relations
and differences among the algorithms.

An important issue in this paper is 
the great flexibility of the idea of
extended ensemble, i.e., an extension is
not necessary restricted to the space
of energy or temperature. In fact,
extensions in the space of arbitrary
macroscopic variables are possible and useful
(Note that some authors already 
noticed this flexibility in the early
stage of the development of extended ensemble
methods, {\it e.g.}, Lyubartsev {\it et al.\/}~\cite{expand}, 
Kerler and Weber~\cite{KW93}. See also
the studies on Adaptive Umbrella 
Sampling algorithm~\cite{M87,HVK92}.).
As we will discuss in Sec~\ref{SS_new},
even unphysical (unrealizable) configurations can be
included in the ensemble, if the fast mixing of the 
Markov chain offsets the increasing cost 
of the sampling procedure.
Such an observation enables us a number of ``special purpose'' 
algorithms, which depend on specific properties of the
model and the computational aims. 

We are also careful
to cross-disciplinary nature of this subject.      
The physicists are no more the only
major users of dynamical Monte Carlo algorithms
and many algorithms have also been developed recently
in various areas that are often overlooked by 
physicists~\cite{MCMC,Neal96,MC}.
For example, Exchange Monte Carlo
(Metropolis-Coupled
Chain, Parallel Tempering) is independently discovered 
by computer scientists working 
for the fifth-generation computer project~\cite{KT90,IHM92},
a statistician~\cite{G91,GT95}, physicists~\cite{H96}, and
the author~\cite{I93}. 

In Sec.~\ref{SS_art}, we discuss the background of
Extended Ensemble Monte Carlo. In the following three sections,
Sec.~\ref{SS_exg}, \ref{SS_temp} and \ref{SS_multi}, three types
of algorithms, i.e., {Exchange Monte
Carlo}, {Simulated Tempering}, 
{Multicanonical Monte Carlo},
are introduced. 
In Sec.~\ref{SSS_rep}, we give
an introduction to Replica Monte Carlo algorithm by Swendsen
and Wang~\cite{SW86,SW88}, 
which interpolates Extended Ensemble Monte Carlo
and Cluster Monte Carlo algorithms. 
Strategies for the construction of 
special-purpose extended ensembles are discussed in 
Sec.~\ref{SS_new}.
In Sec.~\ref{SSS_msse},
extended ensembles with a state space whose dimensionality 
is dynamically varying is discussed. 
In the appendix, 
we compare  three types of extended ensemble algorithms
and discuss their advantages and disadvantages.

This paper was originally written as a part of the
Ph.~D thesis by the author, and then rewritten
as an independent review paper. When I was writing
the manuscript, I discovered several interesting surveys
on this subject. For example,
a lecture note by Marinari~\cite{Mari96}
gives a survey on this field including Exchange Monte Carlo. 
The book~\cite{MC}
provides a cross-disciplinary survey on 
the recent progress of Monte Carlo methods.
Specifically, Berg~\cite{Berg00} in~\cite{MC}
gives a recent review of 
Multicanonical Monte Carlo and related topics. 
A review  on the calculation 
of partition functions (normalizing constants)
by thermodynamic integration and/or
Extended Ensemble Monte Carlo from the viewpoint of statisticians 
is available in~\cite{path}.
Now, there are increasing references
in this field, but I hope that this review gives
fresh perspectives 
both for beginners and experts in this field.

\section{From Natural Ensemble to Artificial Ensemble} \label{SS_art}

Dynamical Monte Carlo algorithms are useful tools
for sampling from  non-Gaussian, highly multivariate
distributions. They, however, often suffer from
slow mixing of the Markov chains, or, in terms of physics, {\it slow
relaxation\/}. Slow relaxation reduces the effective number of samples 
and sometimes leads to wrong results sensitive to
initial states of the Markov chain.
There are several different situations that lead
to slow relaxation: (1) ``{\bf Critical slowing down}'' near second order
phase transition points. (2) ``{\bf Nucleation}'' associated 
with first order phase transitions. (3) {\bf Trapping in metastable states
around local minima} in models with rugged energy landscapes.
Difficulties of the category (3) are often encountered with
``random frustrated systems'' such as models of spin glasses, 
interacting spins in random fields, and heteropolymers.
Slow mixing also appears in complex statistical inference
problems where models (i.e., likelihoods or priors, or both) are 
highly non-Gaussian.

In the period 1985-1995, a powerful strategy
to overcome the difficulties of category (2) and (3),
Extended Ensemble Monte Carlo algorithms, has been
introduced~\footnote{It is believed that they are not useful to
fight against difficulties of 
the category~(1), critical slowing down, to which
Cluster Monte Carlo algorithms~\cite{CMC} and 
Accelerated Hybrid algorithms~\cite{Kron} are successfully
applied.}. 
Simulated Tempering (Expanded Ensembles), 
Exchange Monte Carlo (Metropolis-Coupled Chain, 
Parallel Tempering), Multicanonical Monte Carlo
(Adaptive Umbrella Sampling) are
well-known members of this family. 
While conventional Dynamical Monte Carlo algorithms simulate
a Markov chain whose invariant distribution is a given target distribution
({\it e.g.}, a Gibbs distribution for statistical physics
and a posterior distribution for Bayesian inference),  
Extended Ensemble Monte Carlo algorithms 
sample from artificial ensembles
that are constructed as extensions or compositions of the original
ensembles. 
Fast mixing of the Markov chain in higher temperature (energy, {\it etc.}) components of the artificial ensembles
greatly facilitate the mixing in other components.
Averages over the original
ensemble are calculated by marginalization (in Exchange Monte Carlo),
conditional sampling (in Simulated Tempering), 
or, a {\bf reweighting}  procedure (in Multicanonical Monte Carlo). 
As we will represent in the next section, 
Sec.~\ref{SS_exg}, they can be interpreted as
extensions of Simulated Annealing
algorithms for finite temperature simulations. 

Acceleration of the relaxation
is not the only aim of Extended Ensemble Monte Carlo. 
There are at least two more motivations for
the introduction of artificial ensembles:\\

\noindent {\bf Calculation of integrals or summations}\\
Calculation of multivariate integrals or multiple summations
({\it e.g.}, free energy difference, marginal likelihood difference)
is important in many applications, but cannot be directly
done with conventional  
Dynamical Monte Carlo algorithms. Extended Ensemble Monte Carlo
methods are particularly suitable for the calculations of these quantities.
As we will show in the following sections, 
Exchange Monte Carlo 
naturally gives samples from
a set of distributions
necessary for thermodynamic integration. 
In Multicanonical
Monte Carlo, integrals
are calculated by a reweighting formula.  
In both cases, we can enjoy the advantage 
of fast mixing with extended ensembles 
without additional computational resources.
On the other hand, there are cases where the use of 
an extended ensemble is essential
for the calculation of integrals, as we will discuss 
in the section on Multicanonical Monte Carlo, Sec.~\ref{SS_multi}.\\
  
\noindent {\bf Efficient sampling of ``rare events''} \\
Extended Ensemble Monte Carlo methods are also suitable
for the calculation of the frequencies of  
configurations with small probability in a given ensemble.
For example, we can use them for the calculation of the
free energy surface as a function of one- or two-
macroscopic variables.
The relative error of the computation is
proportional to $1/\sqrt{M}$ where $M$ is the frequency
of independent visits to configurations with a set of values of the 
macroscopic variables. Thus, with a conventional Monte Carlo algorithm,
the histogram in the log-scale contains large noise in low
probability regions. On the other hand, we can compute free energy
surface with much more uniform accuracy with
Extended Ensemble Monte Carlo methods.
We will give some
comments on the implementation of this idea in Sec.~\ref{SS_new}.

\vspace{0,5cm}

Calculation of free energy difference and free energy surface
by artificial ensembles
is especially stressed in the studies on {\bf Expanded 
Ensemble Monte Carlo}~\cite{expand,LFL98,expand96,expandhttp} 
and {\bf Adaptive Umbrella Sampling}~\cite{M87,HVK92,BK97}.
In this context, Extended Ensemble Monte Carlo is 
considered as a descendant of 
{\bf Umbrella Sampling} algorithms
for the calculation of free energy,
which was introduced by Torrie and Valleau~\cite{TV} in 1970s. 
In the original form of Umbrella Sampling algorithms,
artificial ensembles were also used, but systematic
ways for the construction of ensembles had not been
implemented. Implementation of such a procedure
characterizes Extended Ensemble Monte Carlo 
algorithms developed later.

An important feature shared by
Extended Ensemble Monte Carlo
and Umbrella Sampling is that they are {\it not\/} designed for the
direct simulation of {\it natural phenomena\/}. Although 
conventional Dynamical Monte Carlo methods themselves
are something between {\it simulation of physics\/} and
{\it numerical methods\/}, 
they are still strongly motivated
by simulation of physical dynamics -- This is a reason 
why the use of single-spin flip algorithms 
that directly sample Gibbs distributions has 
persisted for a half century.
On the other hand, Extended Ensemble Monte Carlo are free from 
such restrictions. 
It does not mean that insight into the {\it physics}
(or {\it mathematics}, {\it statistics}, \ldots ) of the problems
are unnecessary. On the contrary, they are essential for
the construction of artificial ensembles for
efficient computation. Moreover, the performance of the 
simulation with an extended ensemble
can be regarded as a measure of our understanding of 
the underlying physical phenomena. Our belief is:
\begin{quote}  
{\it If an algorithm based on
a physical picture is efficient, it supports the validity
of the picture; if we understand the physics, we can write
an efficient algorithm.}
\end{quote} 
This manifestation is also applicable to other ``artificial'' 
algorithms, say, Cluster Monte Carlo~\cite{CMC} or 
Hybrid Monte Carlo with acceleration~\cite{Kron}.

\section{Exchange Monte Carlo} \label{SS_exg}

A useful way for the search of ground states of complex models is
{\bf Simulated Annealing} algorithm~\cite{K83,Geman84}. 
The term ``annealing'' indicates
that we start simulations at a high temperature and gradually
decrease temperature to zero. Then, there are more chances of escaping
from shallow local minima and reaching a deep local minimum, or, if we
are lucky enough, attaining the global minimum of the energy function.   
Instead of the inverse temperature $\beta$,
we can use an arbitrary parameter $\lambda$ to interpolate an ``easy'' problem
(a problem with a smooth landscape)
to the original problem (a problem with a rugged landscape). 

Simulated Annealing is, however,
no more than a prescription for {\it optimization}, 
i.e., it is useful for the computation of ground states
of a system but does not exactly 
give finite temperature properties of the system. 
Then, a question arises: {\it Can I extend it for
the sampling from multivariate distributions}, {\it e.g.},
sampling from Gibbs distribution at finite temperatures?
A naive method to achieve 
this purpose is to start the simulation at a high 
temperature and gradually decrease the temperature 
to the target temperature and keep it constant
through the rest of the simulation, where we measure the required 
quantities. This method has, however,  a fatal weak point that 
the annealing is useful for escaping from shallow local minima
but {\it not} for accelerating jumping between 
deep metastable states. To facilitate such jumping, we should
not monotonically decrease the temperature but make it ``{\it up and down}''
alternately. However, at a first glance, any attempt to change the temperature
by external programs, periodic or stochastic, seems to violate the
{\it detailed balance condition}, which is a foundation
of Dynamical Monte Carlo algorithms. 

Here we introduce {\bf Exchange Monte Carlo algorithm}~\cite{H96,H99} 
({\bf Metropolis-Coupled Chain algorithm}~\cite{G91,GT95},
{\bf Time-homogeneous Parallel Annealing}~\cite{KT90}(see also~\cite{IHM92}), 
{\bf Multiple Markov Chain 
algorithm}~\cite{Tesi}, {\bf Parallel Tempering}~\cite{Mari96})
as a solution of the dilemma. The algorithm seems 
to have been independently discovered by
several different groups of authors 
in the period of 1990-1994~\cite{KT90,G91,I93,H96} and, as a result,
has a variety of different names~\footnote{
{\bf J-Walk}~\cite{J-W, MJ-W} algorithm
also uses multiple copies of systems.
In this method, however, the propagation
of configurations is unidirectional, i.e., 
restricted to that from 
a higher temperature to 
a lower temperature. As a result,
the J-Walk algorithm does not exactly 
produce samples from the original distribution, unless
the correlation between samples from the auxiliary simulations at 
a high temperature is negligible (or erased by some off-line manipulation).
Another algorithm closely related to  Exchange Monte Carlo
is {\bf Replica Monte Carlo}~\cite{SW86}
developed by Swendsen and Wang in 1986, 
which we will discuss in Sec.~\ref{SSS_rep}. }.

Consider a set of the distributions 
$\{p_k(\boldx)\}$ with different parameters~\footnote{
A note for the applications to statistical information processing:
The ``parameter'' here often
corresponds to a ``hyperparameter'' when
hierarchical Bayesian models are treated by the algorithm.} 
$\{\lambda_k\}$, $k=1,\ldots,K$ and assume that the parameters are ordered as
$\lambda_1 \geq \lambda_2 \geq \cdots \geq \lambda_K$. 
An example is a family of the
Gibbs distributions defined with inverse temperatures
$\{\lambda_k\}$=$\{\beta_k\}$,
\begin{equation}
\label{Gibbs0}
p_k(\boldx) = \frac{\exp(-\beta_k E(\boldx))}{Z(\beta_k)}
\end{equation}
where $Z(\beta_k)$ is the partition function of the system.
If we denote the variables of $k$th system ($k$th {\it replica}) as 
$\boldx_k$, the simultaneous distribution $\tilde{p}$
of $\{\boldx_k\}$ is written as
\begin{equation}
\label{S}
\tilde{p}(\{\boldx_k\})= \prod_k p_k(\boldx_k) {}_.
\end{equation} 
We introduce two types of update with which
the simultaneous distribution $\tilde{p}$ of eq.~(\ref{S}) 
is invariant. First we consider conventional 
updates in a replica $k$ that satisfy the detailed balance
condition for the corresponding factor $p_k(\boldx_k)$,
{\it e.g.}, local spin flips in a replica. In addition, we
define a {\it replica exchange} between 
replicas which have neighboring 
values of parameters
$\lambda_k$ and $\lambda_{k+1}$.
In this step, candidates of new 
configurations $\widetilde{\boldx}_k$ 
and $\widetilde{\boldx}_{k+1}$ are defined 
by the exchange of configurations of
the replicas: $\widetilde{\boldx}_k=\boldx_{k+1}$ and $\widetilde{\boldx}_{k+1}=\boldx_{k}$. 
If we give the acceptance probability of the
replica exchange flip
by $\max \{1,r\}$ with $r$ defined by
\begin{equation}
\label{r}
r= \frac{p_k(\widetilde{\boldx}_k)p_{k+1}(\widetilde{\boldx}_{k+1})}
{p_k(\boldx_k) p_{k+1}(\boldx_{k+1})} 
=
\frac{p_k(\boldx_{k+1})p_{k+1}(\boldx_{k})}
{p_k(\boldx_k)p_{k+1}(\boldx_{k+1})} 
\, {}_,
\end{equation}
the simultaneous distribution $\tilde{p}$ of eq.~(\ref{S}) 
is invariant under the transition. That is, the detailed
balance condition for the simultaneous distribution $\tilde{p}$
is satisfied.
When $\{p_k(\boldx_k)\}$ is a family of the
Gibbs distributions~(\ref{Gibbs0}) 
defined with inverse temperatures
$\{\beta_k\}$, we can express the ratio 
$r$ as
\begin{equation}
\label{rG}
r = \exp( \, (\beta_k - \beta_{k+1}) \cdot (E({\boldx}_k) 
-E({\boldx}_{k+1})) \,)
\end{equation}

The averages taken over
each factor $p_k(\boldx_k)$ {\it exactly\/} reproduce 
the desired averages at the value $\lambda=\lambda_k$ of
the parameters, because
the transitions defined by the algorithm do not change
the simultaneous distribution eq.~(\ref{S}).  
On the other hand, the states of the replicas
are effectively propagated from high temperatures to
lower temperatures through replica exchanges and 
the mixing of Markov chain is facilitated by the fast relaxation
at higher temperatures (or, in general, at the values of the parameter
$\lambda$ with which the mixing of the Markov chain 
is fast and the entropy
of the distribution is large.). 

A problem is the choice of the points 
$\{\beta_k\}$ or $\{\lambda_k\}$. A naive way is that 
uses a set of points with {\it regular spacing}  
that contains sufficiently high temperatures 
(or, the values of the parameter where the mixing is 
fast and the entropy of the distribution is high).
Of course, we should use the interval 
$|\lambda_{k+1} -\lambda_k|$
that gives a sufficiently large
frequency of replica exchange. 
A more sophisticated way is to allow points of variable spacing
and choose them as the exchange rates are uniform in the prescribed
range of the temperature (parameter). 
Although
the naive method and some tuning {\it by hand} is often enough
for practical applications, it is instructive to see how
we can determine the spacing with this criterion~\cite{H96}
(Discussions on related subjects
from the viewpoint of computational statistics 
are found in~\cite{path,OgawaEguchi}.).
From eq.~(\ref{r}), 
the average $\overline{\log r}$ of the logarithm of the 
ratio $r$ that determines the exchange rate
of the neighboring replicas is
\begin{equation}
\label{rest}
\overline{\log r}=
\sum_{\boldx_k} \sum_{\boldx_{k+1}}
p_k(\boldx_k)p_{k+1}(\boldx_{k+1}) \cdot \log \left \{ 
\frac{p_k(\boldx_{k+1})p_{k+1}(\boldx_{k})}
{p_k(\boldx_k)p_{k+1}(\boldx_{k+1})} 
\right \} \, {}_,
\end{equation}
which is expressed as
\begin{equation}
\label{requal}
\overline{\log r} =
- \left \{ \sum_{\boldx} p_k(\boldx) \log 
\frac{p_k(\boldx)}{p_{k+1}(\boldx)} 
+
\sum_{\boldx} p_{k+1}(\boldx) \log
\frac{p_{k+1}(\boldx)}{p_k(\boldx)} \right \}
 {}_.
\end{equation}
The expression in the braces $\{ \ \}$ 
is  a ``symmetrized'' Kullback-Leibler
divergence $D(p_k||p_{k+1})+D(p_{k+1}||p_k)$ between
$p_k$ and $p_{k+1}$. When $\lambda_k \sim \lambda_{k+1}$ ,
it is approximated by
\begin{equation}
\label{rreduce}
\overline{\log r} \sim
- I(\lambda_k) \cdot (\lambda_{k+1}-\lambda_k)^2
\end{equation} 
with
\begin{equation}
\label{Fisher1}
I(\lambda_k) = - \sum_{\boldx} p_k(\boldx) 
\left.
\frac{\partial^2 \log p_\lambda(\boldx)}{\partial \lambda^2} 
\right
 |_{\lambda=\lambda_k}
= - \left \langle \,\frac{\partial^2 \log p_\lambda(\boldx)}{\partial \lambda^2}
 \, \right \rangle_{\lambda_k} 
\end{equation}
where  $\langle \cdots \rangle_{\lambda_k}$ is average
over the distribution $p_k(\boldx)$. $I(\lambda_k)$ is 
called {\bf Fisher information} in statistics. 
For the case of Gibbs distributions eq.~(\ref{Gibbs0}), 
it is related to 
{\bf susceptibility} $\sigma_E^2 =  
- d\langle E \rangle_\beta/d\beta$ 
to inverse temperature $\beta$ 
\footnote{
The notation $\sigma_E^2$ indicates that it coincides with
the variance of the energy 
$\langle E^2 \rangle_\beta - \langle E \rangle_\beta^2$.}
and {\bf specific heat} $C=(\beta^2/N) \sigma_E^2$ 
{\it per} system size $N$ as
\begin{equation}
\label{Fisher2}
I(\beta_k) =
\left.
\frac{\partial^2 \log Z(\beta)}{\partial \beta^2} 
\right
 |_{\beta=\beta_k}
=
\sigma_E^2 
=
N \cdot \frac{C}{\beta_k^2} {}_.
\end{equation}  
Thus, the interval
$|\lambda_{k+1}-\lambda_k|$ that gives reasonable and uniform
replica exchange rate is given by
\begin{equation}
\label{excon}
I(\lambda_k) \cdot |\lambda_{k+1}-\lambda_k|^2 \sim 1 {}_.
\end{equation}
From the condition eq.~(\ref{excon}), we have
an expression of the density
$Q(\lambda)$ of points $\{\lambda_k\}$ 
\begin{equation}
Q(\lambda) \propto \sqrt{I(\lambda)} {}_,
\end{equation}
which is also written as
\begin{equation}
\label{Qgibbs}
Q(\beta) \propto
\sqrt{\sigma_E^2} = \sqrt{\frac{N \cdot C}{\beta^2}}
\end{equation}
for Gibbs distributions.
The expression eq.~(\ref{Qgibbs}) gives two important 
results. First, it shows that the number of replicas 
that is required for Exchange Monte Carlo increases
with $\sqrt{N}$ when the specific heat is constant. 
It is easy to understand the result when we note 
that the exchange is caused by
fluctuation of the energies of the replicas. 
Another observation from eq.~(\ref{Qgibbs}) is that 
a larger number 
of replicas is required in the region where  
$\sigma_E^2$ or $I(\lambda$) takes larger values,
say, near the critical points of 
second order transitions.

The rest of the problem is how to determine the density
without preliminary knowledge on the specific heat
or Fisher information. The answer~\cite{H99} is that  we can
most easily do it through step-by-step tuning of the number
and/or positions of the points $\{\beta_k\}$ or $\{\lambda_k\}$.
After finishing the tuning process, we perform the simulation
for the sampling of desired quantities with fixed values of all
simulation parameters. This idea is an example of a central 
strategy in Extended Ensemble Monte Carlo algorithms:  
\begin{quote}
{\it Learn (or tune) the optimal value of 
parameters of the algorithm  by a step-by-step 
manner in preliminary runs}.   
\end{quote}
This strategy is more important in Simulated Tempering and 
Multicanonical Monte Carlo discussed in the following sections.

Finally we discuss some concrete results obtained 
by Exchange Monte Carlo and show how it works
in real problems in physics. A field where Exchange Monte 
Carlo is effectively used is studies on spin glasses. 
By using Exchange Monte Carlo, we can explore 
large systems considerably below
$T_c$~\cite{H96,NPR,MMZ01,PCA01}, typically, at 
$T \sim 0.7 T_c$ and system size 
$\sim 16^3$ for 3-dim $\pm J$ Ising spin glass models.
With these calculations, as well as with the use of 
novel methods for the analysis of the data,
we are approaching the nature of the spin-glass phase of
models with short-ranged interaction, which is
a long-standing query in this field. 
Exchange Monte Carlo is also successfully used for
the study of spin glass models with continuous spins, {\it e.g.,}
3-dim Heisenberg spin glass models. Hukushima and Kawamura~\cite{HK99}
reported a strong evidence of
chiral glass transition for the model, 
as well as peculiar properties
of the transition. 
Another, potentially important
field of the application of Exchange Monte Carlo is
simulation of protein models. An application to protein folding is 
already found in Hirosawa~{\it et al.}~\cite{IHM92}~(1992). 
Recent developments and applications to realistic 
peptide models are described 
in~\cite{SO01,Mitsutake01,Hans97,SO99,SO00,SKO00}, 
as well as attempts to combine Exchange Monte Carlo
with Multicanonical Monte Carlo.

\section{Simulated Tempering} \label{SS_temp}

Here we discuss 
{\bf Simulated Tempering algorithm}~\cite{MP92,KR94,IrbackMD}, 
an algorithm
closely related to Exchange Monte Carlo algorithm. 
A similar method called {\bf Expanded
Ensemble}~\cite{expand,expand96,LFL98,expandhttp} 
was introduced by Lyubartsev~{\it et al.} almost at the
same time, whose main aim is the calculation of free energy. 
In this approach, the dilemma of
temperature up-down and detailed balance
is resolved by treating the temperature
itself as a {\it dynamical variable} updated in
Monte Carlo simulations.
That is, we construct Markov chains
whose state vector is a direct product 
{\sf (original states, temperature)}. 
Although the following arguments are easily generalized for an
arbitrary family of distributions $\{p_\lambda(\boldx)\}$ with 
a parameter $\lambda$~\cite{expand}, here and hereafter 
we discuss a family of Gibbs distributions
\begin{equation}
\label{Gibbs}
p(\boldx) = \frac{\exp(-\beta E(\boldx))}{Z(\beta)}
\end{equation}
with a parameter $\beta$, where
$Z(\beta)$ is the partition function.
When we treat the inverse temperature $\beta$ 
as a dynamical variable, the distribution $p(\boldx,\beta)$
in the extended space $(\boldx,\beta)$ is represented as
\begin{equation}
\label{eGibbs}
p(\boldx, \beta) \propto 
\exp(\,-\beta E(\boldx)+g(\beta) \,) {}_.
\end{equation}
Here we have introduced an arbitrary function
$g(\beta)$ of $\beta$ that controls the distribution of $\beta$.
With this definition, it is not difficult to construct 
Markov chains to sample from $p(\boldx, \beta)$.
We simply regard  $-\beta E(\boldx)+g(\beta)$
as the energy of the extended system and simulate it with
ordinary updates of $\boldx$ {\it plus} Metropolis update
of the inverse temperature $\beta$. Here and hereafter, we often 
restrict the value of $\beta$ to discrete values 
$\{\beta_k\}$. With this restriction, the function $g(\beta)$ is
determined by the finite set of the 
values $\{g_k\}=\{g(\beta_k)\}$, $k=1,\ldots,K$. It is a
convenient property for the implementation of the algorithm.
Then, the distribution $p(\boldx, \beta)$ 
in eq.~(\ref{eGibbs}) is represented as
\begin{equation}
\label{eGibbsx}
p(\boldx, \beta_k) =
\frac{\exp(\,-\beta_k E(\boldx)\,)}
{Z(\beta_k)} \cdot \tilde{\pi}_k {}_,
\end{equation}
\begin{equation}
\label{peodo_prior}
\tilde{\pi}_k \propto \exp(\, g_k+
\log Z(\beta_k) \,), \qquad \sum_k \tilde{\pi}_k =1
\end{equation}
Note that the $\log Z(\beta_k)$ term in 
eq.~(\ref{peodo_prior}) comes from the normalization constant 
(partition function) of each component.

If we use the samples of $\boldx$ at a value of $\beta=\beta_k$,
we exactly recover the canonical average at $\beta=\beta_k$,
because $p(\boldx, \beta)$ conditioned on $\beta$
reduces to the original canonical distributions by its definition.
The problem is the choice of the function $g(\beta)$.
Without a proper choice of $g(\beta)$, the value
of $\beta$  gets stuck around an uncontrolled value and
there are {\it no samples} available at the desired
values of $\beta$. 
A naive choice $g_k \equiv 0$  usually
gives unsatisfactory results. 
A reasonable way is to take 
\begin{equation}
g_k=-\log Z(\beta_k) {}_, 
\end{equation}
With this choice, the marginal probability 
$\sum_{\boldx} p(\boldx, \beta_k)=\tilde{\pi}_k$
of $\beta$ is the uniform  distribution on a given set $\{\beta_k\}$
of $\beta$. This means that the temperature varies
in a stochastic way in a prescribed range
and the proportion of the 
time that it stays at a value of $\beta_k$  
is independent of $\beta_k$ in the sufficiently long run.    

But, {\it how can we know the value of $\log Z(\beta)$ prior to
the simulation}?  Any algorithm that requires the 
values of $Z(\beta)$ as inputs appears unrealistic  
because $Z(\beta)$
is usually an unknown 
quantity that is computed through the Monte Carlo
simulation. 
Here we can use the idea of optimizing
the parameters of simulation with the preliminary runs. That is,
the algorithm learns the optimal value of $\{g_k\}$
with the iteration of preliminary runs.
Here we will not discuss the way~\cite{KR94} of tuning further.
Instead we give an account of
similar techniques for multicanonical algorithm
in the next section, Sec.~\ref{SS_multi}.
Note that the optimal {\it spacing\/} of $\{\beta_k\}$ 
can also be estimated in the preliminary runs to enable
sufficiently frequent change of $\beta$.

The analogy with Exchange Monte Carlo algorithm is clear.
The change of parameter(s) $\beta$ (or, in general, $\gamma$)
in Simulated Tempering algorithm corresponds to
the ``replica exchange'' procedure in  Exchange Monte Carlo algorithm.
The mixture distribution 
$\sum_{k=1}^K p(\boldx, \beta_k)$
in Simulated Tempering has a direct correspondence
to the simultaneous distribution eq.~(\ref{S}) in Exchange 
Monte Carlo, when
the set of inverse temperatures of replicas $\{\beta_k\}$ in Exchange Monte Carlo is the same as $\{\beta_k\}$ in Simulated Tempering.   
Formally we can write the correspondence as
\begin{equation}
\label{mixJ}
\sum_{k=1}^K p(\boldx_1, \beta_k) 
\, \Leftrightarrow \, \,
\sum_{\boldx_2, \boldx_3, \cdots, \boldx_K} 
\sum_s \, p_1(\boldx_{P_s1}) \cdot p_2(\boldx_{P_s2})
\cdots p_K(\boldx_{P_sK})
\end{equation}
where $P_s$ is a cyclic shift operator $k \rightarrow \bmod(k-1+s,K)+1$ 
with a shift~$s$.
It is also easy to show that 
the rate of temperature flip in Simulated Tempering is
the same order as the exchange rate in  
Exchange Monte Carlo with the same $\{\beta_k\}$.
In this sense, Exchange Monte Carlo algorithm is
a parallel version of Simulated Tempering  
(thus, called ``Parallel Tempering''
by some authors.). 

In the continuum limit, which is better 
for conceptual arguments,
the mixture distribution $\sum_k p(\boldx, \beta_k)$ 
in eq.~(\ref{mixJ}) is represented as
\footnote{
The notation $\tilde{\pi}(\beta)$ is motivated by the analogy to
Bayesian statistics. In fact, the distribution $\tilde{\pi}(\beta)$ 
is at times called a ``pseudo prior'' by 
statisticians~\cite{GT95}.
It is formally regarded as
a prior distribution for a (hyper)parameter $\beta$, 
but determined for the convenience of the 
computation.}
\begin{equation}
\int p(\boldx, \beta) \, d\beta = \int
\frac{\exp(\,-\beta E(\boldx) \,)}{Z(\beta)} \cdot \tilde{\pi}(\beta)
d \beta
\end{equation}
where
\begin{equation}
\label{pseudo_prior_c}
\tilde{\pi}(\beta) \propto \exp(\, g(\beta)+
\log Z(\beta) +Q(\beta) \,)  {}_.
\end{equation}
Here $Q(\beta)$ represent the relative number of 
the points $\{\beta_k\}$ between $\beta < \beta_k < \beta+d\beta$
in both of Exchange Monte Carlo and 
Simulated Tempering. Consider the case
$g(\beta)=-\log Z(\beta)$ in eq.~(\ref{pseudo_prior_c}).
If we use $Q(\beta) \propto \sqrt{I(\beta)}$ that gives the uniform exchange
rate for Exchange Monte Carlo and the uniform
{\it temperature flip} rate for Simulated Tempering, 
the mixing distribution $\tilde{\pi}(\beta)$
coincide with
{\bf Jeffreys' prior}~\cite{Jeffreys}:
\begin{equation}
\int p(\boldx, \beta) d\beta \propto \int
\frac{\exp(\,-\beta E(\boldx) \,)}{Z(\beta)} \cdot \sqrt{I(\beta)}
d \beta {}_.
\end{equation}

Simulated Tempering and related methods 
are successfully used
for various problems in physics and statistics. 
For example, a random field Ising model on a $10^3$ lattice
at low temperatures is treated in the original paper 
of Marinari and Parisi~\cite{MP92}. 
Studies~\cite{WM94,expand96,LFL98,EP95,EP96,WYNP00} based on the idea 
of expanded ensembles~\cite{expand} 
will also be discussed in Sec.~\ref{SS_new} 
and Sec.~\ref{SSS_msse}.
Geyer and Thompson~\cite{GT95} discussed an application 
to statistical inference on propagation of genes of rare
recessive disease on a pedigree. The problem is computation
of probability distributions of career status 
of genes over a large pedigree from observed data, for which
conventional Dynamical Monte Carlo
suffers from slow mixing and non-ergodicity of dynamics.
By using a version of Simulated Tempering 
(see Sec.~\ref{SS_new} of this survey), 
they successfully treated problems that contain thousands
of individuals. 

In most situations, however, Exchange Monte Carlo
is more convenient than Simulated Tempering. 
Some of the advantages of Exchange Monte Carlo   
are discussed in Sec.~\ref{SS_appendix}.
An exception occurs when the dimension of the system
is so large that it is impossible to
store a number of replicas
in the memory of our computer. In this case,
Simulated Tempering has an obvious advantage.
Simulated Tempering may also be useful
for the conceptual studies on extended ensembles.

\section{Multicanonical Monte Carlo} \label{SS_multi}

The third method is {\bf Multicanonical Monte Carlo} 
algorithm~\cite{BergN,BergC,Smith,Berg00}. Methods based on a 
similar idea are also known as {\bf Adaptive Umbrella
Sampling} algorithm~\cite{M87,HVK92,BK97}. 
In their original forms,  Multicanonical Monte Carlo
deals with extensions in the space of the energy, while
Adaptive Umbrella Sampling focused on the extensions 
in the space of a reaction coordinate. Now, they are being 
merged and we can freely construct algorithms optimal for
our purpose, which will be discussed in Sec.~\ref{SS_new}.

Unlike the algorithms discussed in the previous sections,
multicanonical algorithm deals only with 
an exponential family of distributions. 
First, we discuss the case
of the family of Gibbs distributions 
eq.~(\ref{Gibbs}) with different inverse temperatures $\beta$.
The {\it density of state} $D(E)$ on the energy axis 
is defined so that the number of the state 
$\boldx$ satisfying $E < E(\boldx) <E+dE$ is $D(E)dE$.
The partition function (the normalization constant)
at inverse temperature $\beta$ is written as
\begin{equation}
Z(\beta)= \int \exp(-\beta E) D(E) dE{}_.
\end{equation}
{\it Multicanonical algorithm is defined as Dynamical Monte Carlo
sampling with the weight proportional to $D(E(\boldx))^{-1}$ } instead of
the original canonical weight $\exp(-\beta E(\boldx))$.
The distribution defined with this weight is called 
{\bf multicanonical ensemble}. 
With the definition of $D(E)$ and 
$D(E) \cdot D(E)^{-1}=1$,
it is easy to see that 
the marginal distribution of $E$ becomes constant
within the region where $D(E) \neq 0$, i.e.,
the energy $E$ of the system takes the
values in $E \sim E+dE$ with an equal
chance in a long simulation
\footnote{Note that $D(E)$ is a severely varying function
of a macroscopic variable $E$ and the choice of $D(E(\boldx))^{-1}$
as a weight severely penalized the appearance of the states $\boldx$
with a large value of $D(E(\boldx))$, which are usually nearly random
 ``high temperature'' configuration. Multicanonical sampling
corresponds to random selection
of $\boldx$ with the value of $E$ after the random 
sampling of the energy $E$. The point is that it is very different from
random sampling of $\boldx$, which gives almost surely 
a sample of large $E$. }.  
This results in a random
walk in energy space. It is similar to the random walk
on temperature axis in Simulated Tempering, but there is no
explicit temperature variable in Multicanonical Monte Carlo.
When we introduce {\bf microcanonical
ensemble} with an energy $E_0$ defined by
\begin{equation} \label{micro}
p_{E_0}(\boldx) = \frac{\delta(E(\boldx)-E_0)}{D(E_0)} {}_,
\end{equation}
multicanonical ensemble $p_{mul}(\boldx)\propto 1/D(E(\boldx))$ is considered as 
the mixture 
\begin{equation}
p_{mul}(\boldx) = \int dE_0 \,\, p_{E_0}(\boldx) \, \tilde{\pi}(E_0) 
\end{equation}
of microcanonical ensembles 
with the uniform pseudo prior $\tilde{\pi}(E_0)= const.$ for the energy.
In this sense,  multicanonical ensemble is 
a special type of Expanded Ensemble~\cite{expand}, whose
components are microcanonical distributions.

{\it How can we recover the 
canonical averages from the simulation 
with the weight $D(E(\boldx))^{-1}$}?
 A {\bf reweighting formula}
\begin{equation}
\label{reweight} \label{reweightT}
\langle A(\boldx) \rangle_\beta = 
\frac{
\sum_j A(\boldx^j) \cdot \exp(-\beta E(\boldx^j)) \cdot D(E(\boldx^j)) 
}
{
\sum_j \exp(-\beta E(\boldx^j)) \cdot D(E(\boldx^j)) 
}
\end{equation} 
gives an answer, which gives a method for the
reconstruction of the canonical 
average $\langle A(\boldx) \rangle_\beta$
at $\beta$ of an arbitrary
quantity $A(\boldx)$. Here
the summation $\sum_j$ is taken
over the samples $\{\boldx^j\}$ 
generated by the simulation of a Markov chain
whose invariant densities is $D(E(\boldx))^{-1}$.
The eq.~(\ref{reweight}) means that each observation $A(\boldx^j)$ is
multiplied by the factor $D(E(\boldx^j))$ that 
cancels the weight $D^{-1}(E(\boldx^j))$ used in the simulation and
reweighted by the canonical weight 
$\exp(-\beta E(\boldx^j))$. We can also introduce a reweighting formula 
for the normalization constant (partition function) $Z(\beta)$ as
\begin{equation}
\label{reweightZ} \label{reweightZT}
\frac{Z(\beta)}{V}= 
\frac{
\sum_j \exp(-\beta E(\boldx^j)) \cdot D(E(\boldx^{j}))
}
{
\sum_j D(E(\boldx^{j}))
}
\end{equation}
where $V$ equals to the total volume of the configuration space or 
the total number of the configurations ({\it e.g.}, $2^N$ for $N$ binary
variables).
For earlier studies on {\it reweighting\/} in Dynamical Monte Carlo methods,
see~\cite{FS88,FS89,SWFrev}. 

Now we will discuss an essential part of the algorithm:
{\it How to sample with the weight $D(E(\boldx))^{-1}$ 
without prior knowledge on $D(E)$.} This is a basic problem, because
$D(E)$ is the kind of the quantities which we want to calculate
by the simulation, just as $Z(\beta)$ in Simulated
Tempering algorithm. The answer is, again, step-by-step learning
procedure, which we will discuss in detail here.
Note that we do not need to know the value $D(E)$ exactly.
An approximation $\tilde{D}(E)$ to $D(E)$ in a region 
$E_{min} < E < E_{max}$, with which we can safely
apply the reweighting 
procedure eq.~(\ref{reweight}) 
and eq.~(\ref{reweightZ})
is enough for our purpose.
Note also that
multiplication of a constant factor to $\tilde{D}(E)$ does not
change the result.
Paying an attention to these remarks, 
we introduce an iterative procedure
({\bf preliminary run}s)
to approximate $D(E)$, starting from
$D^{0} \equiv const.$ (or some initial guess).
Here and hereafter we assume that
the energy $E$ takes discrete values $\{E_k\}, \,(k=1,\ldots,K)$, 
and $\tilde{D}(E)$ are 
represented by the values
$\{\tilde{D}_k\}=\{\tilde{D}(E_k)\}$.
\begin{enumerate}
\item Simulation and Histogram Construction:
Simulate a Markov chain 
with the weight $D^t(E(\boldx))^{-1}$ and 
record the frequencies $h^t_k$ that the energy $E$
takes the value $E_k$ for each $k$ ({\bf Histogram Construction}).
Here, we can use arbitrary dynamics with which
the density ${D^t(E(\boldx))}^{-1}$ is invariant.

\item
Update the weight : Define new values of the weight by
\begin{equation}
\frac{1}{D^{t+1}_k} : = \frac{1}{D^t_k} \cdot  
\frac{1}{h^t_k+\epsilon} {}_,
\end{equation}
or, equivalently,
\begin{equation}
\log D^{t+1}_k  : = \log D^t_k +  
\log (h^t_k+\epsilon) {}_.
\end{equation}
Here $\epsilon$ is a constant, which is added to remove
the divergence with $h^t_k=0$ (i.e., no visit to $E=E_k$). 
For example, we can use $\epsilon=1$.

\item
Normalization :
\begin{equation}
\log D^{t+1}_k : = \log D^{t+1}_k - \frac{1}{K} \sum_k \log D^{t+1}_k
\end{equation}
This normalization procedure is added for the convenience 
of monitoring convergence and not essential
(Adding a constant to all $\log D^{t+1}_k$
does not change the simulation.). 

\item 
Set t:=t+1.

\end{enumerate}
After we find an appreciate weight $\{\tilde{D}_k\}$ with
the iteration of the preliminary runs, a {\bf measurement run}
({\bf production run})
is performed. In the measurement run, 
we collect samples with a fixed $\{\tilde{D}_k\}$
and apply reweighting formulae
eq.~(\ref{reweight}) and eq.~(\ref{reweightZ}) 
for the calculation of canonical averages,
where $D(E_k)$ is substituted 
for its approximation $\tilde{D}_k$.

This simple iterative procedure, 
sometimes referred to as (the learning stage of) 
the {\bf entropic sampling} method~\cite{Lee} is
sufficient for many practical problems. 
When the system is very large
or has a continuous energy spectrum, we 
should replace histogram construction by a more sophisticated
density estimation method. Another problem of
the above-mentioned procedure is that it is sensitive to the fluctuation
of frequencies of visits to $E_k$. 
Some authors proposed
estimators of $\tilde{D}_k$ 
based on the ratio of $h^t_k/h^t_{k+1}$ of the neighboring frequencies
(or the ratio of the frequencies $E_k \rightarrow E_{k+1}$ and
$E_{k+1} \rightarrow E_k$  of the transition~\cite{Smith,SB96})
for the improvement of the performance in the tuning 
stage.  ``Flat Histogram Monte Carlo'' 
(see~\cite{Wang00,Wang01} and references therein)
can also be considered as an efficient way 
to realize Multicanonical Ensemble,
although it has a different origin~\cite{Oliv} and 
its own perspective.  

When we deals with systems with quenched disorders, we usually do not
have prior knowledge of the upper and lower bounds of $E$. In such cases,
we apply the iterative procedure assuming a sufficiently wide region 
of $E_{min} < E <E_{max}$. Then, we conclude there is no
energy level at $E_k$, if $h_k=0$ even with a sufficiently  
large value of the weight $1/D^t_k$ and in a
long run of the simulation. 
In principle, we can neglect regions that
do not contribute required canonical averages with eq.~(\ref{reweight}),
but should be careful to include a sufficiently high energy region
(or, in general, a high entropy region) to facilitate the relaxation.
Determination of lower energy bound (as well as
higher energy bound if entropy is also small there) is often the 
most time-consuming part of the algorithm. If the length of 
each run of preliminary  simulation
is not sufficient, 
we often observe ``oscillation'' of the histogram at the extremes
of the energy band, i.e., $h_k$ that takes a small value 
in $t$th simulation becomes large in $(t+1)$th simulation, 
and, again become small in $(t+2)$th step $\ldots$ and it does 
not converge.

We emphasis that a single simulation with an approximately 
multicanonical weight $1/\tilde{D}(E(\boldx))$ is enough
to obtain canonical averages at any $\beta$. 
It is because 
a random walk with the weight $1/\tilde{D}(E(\boldx))$ 
covers whole range of the energy and we can collect
information at all possible values of energy using them.
An examination of the reweighting formulae also
shows that the efficiency of the 
reweighting procedure relies on the flatness of
the marginal distribution of the energy
in the range of $E_{min} < E < E_{max}$. The flatness
ensures that the weight $\exp(-\beta E) \tilde{D}(E)$
of a sample in reweighting formula~(\ref{reweightT}) 
is a Gaussian-like distribution 
with the width $\propto 1/\sqrt{N}$.
Thus, the number of the samples that contribute to a canonical
average is proportional to $1/\sqrt{N}$ for any value of $\beta$.
For the purpose of a comparison,
consider the sampling with
the canonical weight $\exp(-\beta^\prime 
E(\boldx))$ with a temperature $1/\beta^\prime$ and reweighting to 
the temperature $1/\beta$ ($\beta>\beta^\prime$)~\cite{FS88}. 
The corresponding reweighting formula
\begin{equation}
\label{reweightX}
\langle A(\boldx) \rangle_\beta = 
\frac{
\sum_j A(\boldx^j) \exp((\beta^\prime-\beta)\cdot E(\boldx^j)) 
}
{
\sum_j \exp((\beta^\prime-\beta) \cdot E(\boldx^j)) 
}
\end{equation} 
is formally valid for any pair $\beta^\prime$ and $\beta$, but
not useful except when $\beta^\prime$ are close to $\beta$.
In this case, the marginal distribution of the energy of samples 
is a Gaussian-like distribution with the width $\propto 1/\sqrt{N}$,
while the factor $\exp((\beta^\prime-\beta)\cdot E)$ is quickly
decreasing function of $E$ with a decay constant $\propto N$,  
As a result, an
exponentially small fraction of samples
contribute to the required average for a large system size $N$.

So far we discussed multicanonical algorithm with Gibbs distributions.
It is easy to extend it to a one-parameter
exponential family
\begin{equation} \label{expfa1}
p_\lambda(\boldx)= 
\frac{\exp( \beta \cdot E(\boldx) +\lambda \cdot f(\boldx))}{Z(\lambda)}
\end{equation}
parameterized by $\lambda$. We can choose 
any physical quantity as a function $f(\boldx)$, 
{\it e.g.}, volume, magnetization, dihedral angle, radius of gyration, 
and replica overlap. To define the algorithm, we use $D(f(\boldx))$ instead of 
$D(E(\boldx))$, where $D(f)df$ is defined as the number of the
states that satisfy $f < f(\boldx) < f+df$.
We can also consider multi-dimensional extensions for a
multi-parameter exponential family~\cite{SLV97,BK97,HIGO97,
NN98,iba98,Chiken99,Hatano},
\begin{equation} \label{expfa2}
p_{\boldlam}(\boldx)= \frac
{\exp (\sum_l \lambda_l \cdot f_l(\boldx) ) }
{Z(\boldlam)}
\end{equation}
where $Z(\boldlam)=Z(\lambda_1,\lambda_2,\ldots,\lambda_L)$ 
is the partition function. In this case,
we define a simultaneous density 
$D(f_1,f_2,\ldots,f_L)$ and estimate it by multivariate histogram
construction with preliminary runs. Of course, we can treat 
$\beta$ as one of the parameters $\lambda_l$. 
It seems, however, not possible to 
extend multicanonical algorithm
beyond exponential family
that are determined by a set of sufficient statistics.
In terms of physics, multicanonical ensemble is defined
using extensive quantities conjugated
to an ``external force'' and cannot be generalized
to the cases to which we cannot specify such quantities.   
At this point, it differs from Exchange Monte Carlo or Simulated
Tempering, which can apply to any family of distributions
$p_{\boldlam}(\boldx)$
\footnote{
An example of distributions 
that is not part of exponential family 
is Cauchy distributions
$p_\lambda (x)= (\lambda/\pi) \cdot 1/(x^2+\lambda^2)$
with a scale parameter $\lambda$.
}.

On the other hand, {\it there are cases that are well 
treated by Multicanonical Monte Carlo, but not by the other
two methods.} Typical examples are provided by 
models with first order
phase transition with latent heat. In these cases, there
is a region on the energy axis that cannot be covered by
a Gibbs distribution, i.e., for any value of inverse 
temperature $\beta$, there is negligible 
chance of finding a sample $\boldx$ with a value
of $E(\boldx)$ in the region. In the case of multicanonical
Monte Carlo, the ensemble
constructed by the iterative learning procedure
contains samples with these missing values of the energy, which make
the algorithm work as we expect. On the other hand, Exchange Monte Carlo
and Simulated Tempering do not work in these cases, because any
mixture of canonical ensembles (Gibbs distributions) contains
little portion of samples with the energy in the gap region.  
In a physical interpretation that applies to models of liquids
and lattice spin models,  multicanonical ``energy''
$-\log D(E(\boldx))$ contains an artificial correction term to
the interfacial tension of a droplet that 
makes the critical radius of a droplet zero
and also controls the growth of a droplet after nucleation. 
While it seems
very difficult to design such a term by hand, multicanonical
algorithm automatically learns a term with desired properties
with a histogram construction 
(or, some alternative iterative tuning 
methods.)~\footnote{In some cases, we need additional techniques 
to estimate the weights for multicanonical calculation
in preliminary runs.
In the work~\cite{BergN} of Berg and Neuhaus
that deals with a Potts model, approximate weights
found in smaller systems is used as an initial guess 
of the corresponding weights in larger systems.}.

When there is no first order transition and resultant 
``phase coexistence regions'',
{\it how can we relate a multicanonical ensemble to the mixture of
Gibbs distributions?} At first sight,
it may be natural to expect that the mixture
with $\tilde{\pi}(\lambda) \propto \sqrt{I(\lambda)}$ (Jeffreys' mixture)
approximates well the corresponding multicanonical ensembles. It is,
however, {\it not} true. In fact, the choice $\tilde{\pi}(\lambda) \propto
I(\lambda)$, which gives larger 
weight to high specific heat regions, provides a better
approximation to the multicanonical ensemble. 
We illustrate the result with
a simple example of binomial distribution
\begin{equation}
p_p(n)={}_NC_n \cdot p^n (1-p)^{N-n} {}_,
\end{equation}
which is expressed as an exponential family 
\begin{equation}
p_\lambda(n)=\frac{\exp(\lambda \cdot n)}{Z(\lambda)}, \qquad 
\lambda= \log \frac{p}{1-p}
\end{equation}
with a parameter $\lambda$.
For this model, 
$Z(\lambda)= 1/{}_NC_n \cdot 1/(1-p)^N$,
$I(\lambda)=N p(1-p)$, $I(p)=N/(p(1-p))$ and
$d\lambda/dp = 1/(p(1-p))$. By using them, it is
easy to show that the Jeffreys' prior $\tilde{\pi}(\lambda)$ 
of $\lambda$ is given by
\begin{equation}
\tilde{\pi}(\lambda)d\lambda \propto
\sqrt{I(\lambda)}d\lambda = \sqrt{I(p)}dp = \frac{\sqrt{N} \, 
dp}{\sqrt{p(1-p)}} {}.
\end{equation}
On the other hand, the mixture with
\begin{equation}
\label{unifp}
\tilde{\pi}(\lambda)d\lambda \propto
I(\lambda)d\lambda = N dp
\end{equation}
gives the uniform density
on $n$ axis, i.e., it is a multicanonical ensemble. 
It is easily confirmed with the identity
\begin{equation}
\int p_p(n) dp = \int {}_NC_n \cdot p^n (1-p)^{N-n} 
\, dp = \frac{1}{N+1} {}_,
\end{equation}
whose right-hand side does not contain $n$.

Let us discuss some of the typical results obtained by 
Multicanonical Monte Carlo. One of the attractive
applications is found in the field of protein 
folding~\cite{HO,HO2,Mitsutake01}.
An illustrative example of the ability of avoiding
local optima by Multicanonical Monte Carlo is shown
in Fig.~2 and Fig.~3 of~\cite{Mitsutake01}, where Multicanonical 
Monte Carlo efficiently realize $\alpha$-helical 
conformations expected by
laboratory experiments while conventional
Monte Carlo fails.  An advantage of Extended Ensemble Monte Carlo
is, however, more clear in the examples where fluctuations
among the structures are significant. Such examples are also
found in literatures, for example, \cite{Mitsutake01} and~\cite{HGON01}. 
In~\cite{HGON01}, a $\beta$-hairpin peptide of $16$ residue
in explicit water (139 peptide atoms and 1060 water molecules)
is simulated, and it is shown that 
the molecule fluctuates around conformations
classified into several clusters at a physiological temperature.

As we have discussed in this section, an advantage of
Multicanonical Monte Carlo over other Extended Ensemble
Monte Carlo algorithms is that it enables the treatment of 
first-order transitions. In this respects, the
paper~\cite{BergN} by Berg and Neuhaus already gave 
an impressive example,
i.e., simulation of 10-state Potts model up to the size $100^2$. 
Multicanonical Monte Carlo and its variants are also useful
for the simulation of liquids and gas~\cite{WildingR}
(see also Sec.~\ref{SSS_msse} of this paper), where we encounter 
classical examples of first-order transitions.

\section{Replica Monte Carlo} \label{SSS_rep}

Replica Monte Carlo algorithm~\footnote{
Note that a similar term ``{\bf Replica Exchange Monte Carlo}'' is sometimes 
used as a synonym of ``Exchange Monte Carlo''.} 
by Swendsen and Wang~\cite{SW86,SW88}
is one of the pioneering works on 
Dynamical Monte Carlo algorithms 
that use multiple copies of the system.
In fact, it includes Exchange Monte Carlo algorithm
as a limit and precedes  
any study on Exchange Monte Carlo 
referred in this paper. However, {\it cluster dynamical
aspect\/} of the algorithm is highly stressed
in the original representation~\cite{SW86} and it seems
not trivial to extract Exchange Monte Carlo algorithm from it. 
Cluster identification using a pair of replicas
is a really ingenious and attractive idea
itself, but it severely restricts 
the application of the algorithm when we persist in it.

In this section, we give an introduction
to cluster dynamics of Replica Monte 
Carlo algorithm and clarify the relation 
between Replica Monte Carlo 
and Exchange Monte Carlo.
It seems that there has been no concrete attempt to generalize
the idea of cluster dynamics in Replica Monte Carlo
beyond Ising models, although some suggestions
are given in~\cite{SW86}. Thus, we restrict our attention  
to Ising models with inhomogeneous couplings $\{J_{ij}\}$. 
For this class of models, the Gibbs distribution is written as
\begin{equation}
\label{GibbsRMC}
p(\{x_i\})= \frac{\exp(\beta \sum_{(ij)} J_{ij} x_ix_j)}{Z(\beta)}
\end{equation} 
where $x_i \in \{\pm 1\}$ is a Ising spin variable
that is defined on the vertex $i$ of a graph $G$ ({\it e.g.},
a square lattice) and
$Z(\beta)$ is the partition function. The summation $\sum_{(ij)}$ runs
over the all pairs $(ij)$ where $i$ and $j$ are neighboring on $G$,
i.e., the edge $(i,j) \in G$.   
Consider a set of the
Gibbs distributions eq.~(\ref{GibbsRMC}) defined with temperatures
$\{\beta_k\}$. We denote the spin variables of $k$th system as 
$\{x_i\}^k$. Then, the simultaneous distribution $\tilde{p}$
of $\{ \{x_i\}^k, \ k=1,\ldots,K\}$ is written as
\begin{equation}
\label{rep}
p_k(\{x_i\}^k)= \frac{\exp(\beta_k \sum_{(ij)} J_{ij} x_i^k x_j^k)}{Z(\beta_k)}
\end{equation} 
\begin{equation}
\label{SRMC}
\tilde{p}(\{\{\boldx_i\}^k\})= \prod_k p_k(\{x_i\}^k)
\end{equation} 
where $Z(\beta_k)$ is the partition function of the $k$th system ($k$th replica).

So far, the construction is the same as the one
for Exchange Monte Carlo.
In Replica Monte Carlo by
Swendsen and Wang, 
non-local {\it cluster update\/}
is used as well as usual single-spin flip update in 
each replica.
It is defined for a pair
of replicas which have neighboring 
values of temperatures $\beta_k$ and $\beta_{k+1}$.
To define clusters in a pair of configurations 
$\{x_i\}^k$ and $\{x_i\}^{k+1}$, 
we introduce variables $t_i = x_i^k x_i^{k+1}$
and define an equivalence relation $\equiv$ among the vertices of $G$:
If $(i,j) \in G$ and $t_i=t_j$ then $i \equiv j$.  
We define {\it clusters} as the equivalence classes with the relation $\equiv$. 
Note that there are two types of clusters
distinguished by the values of $t_i$.
In this paper, we call them {\it parallel clusters} ($t_i=1$) and {\it anti-parallel clusters} ($t_i=-1$), respectively.
Then, we define {\it cluster flip} as simultaneous flips of spins
in a cluster in both replicas. That is, we choose a cluster $c$ defined
with configurations $\{x_i\}^k$, $\{x_i\}^{k+1}$ and
generate candidates of new configurations ${\{\widetilde{x}_i\}^k}$ 
and $\{\widetilde{x}_i\}^{k+1}$ defined by: ${\widetilde{x}_i^k}
=-x_i^k$ if $i \in c$
else ${\widetilde{x}_i^k}=x_i^k$; \ ${\widetilde{x}_i^{k+1}}=
-x_i^{k+1}$ if $i \in c$ else ${\widetilde{x}_i^{k+1}}=x_i^{k+1}$.
With this ${\{\widetilde{x}_i\}^k}$ 
and $\{\widetilde{x}_i\}^{k+1}$, the acceptance probability of the
cluster flip is given by $\max \{1,r\}$ where
\begin{equation}
r= \frac{p_k(\{\widetilde{x}_i\}^k)p_{k+1}(\{\widetilde{x}_i\}^{k+1})}
{p_k(\{x_i\}^k) p_{k+1}(\{x_i\}^{k+1})} \, {}_.
\end{equation}
When we define the boundary $\partial c$ of a cluster $c$ 
as the set of edges $(i,j)$ that satisfy $j \in c$ and $i \not\in c$,
the logarithm of the ratio $r$ is expressed as
\begin{equation}
\label{rrsw}
\log r = -2 \cdot (\beta_k - \beta_{k+1}) \cdot
\sum_{(i,j) \in \partial c} 
J_{ij} x^k_i x^k_j \, {}_.
\end{equation}
This expression gives $r \sim 1$ for $\beta_k \sim \beta_{k+1}$,
i.e., a cluster flip is accepted with a high probability for a pair
of replicas with sufficiently close temperatures, even when
$|\beta_k J_{ij}|$ are large. These cluster flips
share some characteristics with {\bf crossover} in {\bf Genetic algorithm}s   
in the sense that {\it they generate a new configuration from 
two existing configurations}. However, in contrast to random 
crossover that is rarely accepted with
a large cluster exchange, cluster flips in Replica Monte Carlo 
are designed to realize high acceptance ratio while satisfying 
a detailed balance condition.

Let us discuss connections to Exchange Monte Carlo algorithm
introduced in Sec.~\ref{SS_exg}.
In the case of Exchange Monte Carlo, we also consider the
simultaneous distribution $\tilde{p}$ of eq.~(\ref{S}) and define
{\it replica exchange} between 
replicas which have neighboring 
values of temperatures $\beta_k$ and $\beta_{k+1}$.
The identification of clusters is, 
however, not necessary in Exchange Monte Carlo.
Instead, we define candidates of new 
configurations ${\{\widetilde{x}_i\}^k}$ 
and $\{\widetilde{x}_i\}^{k+1}$ by the 
exchange of configurations of
replicas: For all $i$, 
${\widetilde{x}_i^k}=x_i^{k+1}$ and ${\widetilde{x}_i^{k+1}}=x_i^{k}$. 
Then, the acceptance probability of the
cluster flip is given by $\max \{1,r\}$ with $r$ defined
by eq.~(\ref{r}). The explicit form of $\log r$ for the present model 
is given by
\begin{equation}
\label{exg}
\log r = -2 \cdot (\beta_k - \beta_{k+1}) \cdot
\sum_{(i,j)} 
J_{ij} \cdot ( x^k_i x^k_j -  x^{k+1}_i x^{k+1}_j    )\, {}_.
\end{equation}
Although the implementation of Exchange Monte Carlo does not
require the cluster identification procedure, it is useful
for our purpose to rewrite it with the language of the clusters defined 
in Replica Monte Carlo. 
We define the sets $c^+=\bigcup_{m \in M_{+}} 
c_m$ and $c^-=\bigcup_{m \in M_{-}} c_m$ 
as joint sets of parallel and anti-parallel 
clusters, respectively. 
Here $M_{\pm}$ indicates the set of 
indices of parallel/anti-parallel clusters. 
Then the {\it replica exchange} is equivalent to {\it flipping of
the joint of all anti-parallel clusters $c^-$ }:
${\widetilde{x}_i^k}
=-x_i^k$ if $i \in c^-$
else ${\widetilde{x}_i^k}=x_i^k$; \ ${\widetilde{x}_i^{k+1}}=
-x_i^{k+1}$ if $i \in c^-$ else ${\widetilde{x}_i^{k+1}}=x_i^{k+1}$.
Thus, eq.~(\ref{exg}) is written as   
\begin{equation}
\label{rexg}
\log r = -2 \cdot (\beta_k - \beta_{k+1}) \cdot
\sum_{(i,j) \in \partial c^-} 
J_{ij} x^k_i x^k_j \, {}_.
\end{equation}
where the boundary $\partial c^\pm$ 
of $c^\pm$ is defined as the sets $(i,j)$ that $i \in c^\pm$ and 
$j \not\in c^\pm$ (With this definition $\partial c^+ = \partial c^- \, = \,
\bigcup_{m \in M_{+}}  \partial c_m 
\, = \, \bigcup_{m \in M_{-}}  \partial c_m$.). 
Note that the flip of the set $c^+$ of
all parallel clusters is essentially
equivalent to the flip of $c^-$ for an Ising model with
up-down symmetry. It reduces
to the exchange of the replicas when the flip of the all spins
in replicas ${\widetilde{x}_i^k}=-\widetilde{x}_i^{k}$ 
and ${\widetilde{x}_i^{k+1}}= - \widetilde{x}_i^{k+1}$ is added
after the cluster flip.  

The analogy between eq.~(\ref{rrsw}) and eq.~(\ref{rexg})
is obvious. If there are only two clusters, one is 
parallel and the other is anti-parallel, both algorithms
give essentially the same dynamics. In this case, Exchange Monte Carlo
has an advantage, because it does not require a cluster
identification procedure. On the other hand,
with this observation, it is not difficult
to construct a family of algorithms     
that interpolate Replica Monte Carlo and
Exchange Monte Carlo~\footnote{In the original 
paper~\cite{SW86},
there is no specification on the dynamics
for cluster flips in Replica Monte Carlo. 
In this sense, Replica Monte Carlo virtually contains these interpolations.
However, there seems no explicit
suggestion on naive multi-spin flips
in the references of Replica Monte Carlo (a comment on the use of percolation
representation for cluster update is found in~\cite{SW88}).}.
In these algorithms, we construct clusters just the same way as that in
Replica Monte Carlo, but flip more than one cluster simultaneously with the
restriction of {\it only the same types of clusters can be flipped
at one time}. That is, we generate a new candidate of configurations
by the flip of the union of a set of parallel clusters, or,
a set of anti-parallel clusters.  It is easy to see
that Replica Monte Carlo and 
Exchange Monte Carlo is regarded as two extremes of the algorithm,
where the number of clusters updated in a single 
trial is {\it one} and {\it maximum} respectively. Again, there
is a discontinuity in the performance between 
Exchange Monte Carlo and the Exchange Monte Carlo-like limit 
of generalized Replica Monte Carlo, which is caused by 
the cost of the cluster
identification procedure.

A weakness of the cluster identification procedure 
in Replica Monte Carlo is that it is not easy to generalize
it for complicated models, {\it e.g.}, lattice or off-lattice 
protein models.
Another weak point might be found in the way of 
defining clusters itself.
In Replica Monte Carlo, a cluster is defined 
with a pair of replicas that is {\it mutually
independent}, possibly coming from very different regions 
of the phase space. Whether the clusters identified by such a way
have adequate properties will depend on the model to be examined.
On the other hand, Exchange Monte Carlo algorithm
without cluster dynamics is much more robust
and has been applied to a variety of models 
in various fields.

In the literature, two-dimensional $\pm J$
Ising spin glass models seem the only example 
with which the efficiency of Replica Monte Carlo 
is quantitatively analyzed~\cite{SW86,H01}.
A recent study by Houdayer~\cite{H01} 
treated the model on $12^2 \sim 100^2$ lattices by a modification
of Replica Monte Carlo~\footnote{Houdayer's algorithm uses 
multiple series of replicas with the same set of temperatures. 
Cluster dynamics is defined only with pairs 
of replicas with same temperatures, while Exchange Monte Carlo
is applied to replicas with different temperatures.}
The paper reports that the algorithm performs much better than
Exchange Monte Carlo for the model and
thermal equilibrium is attained even 
at very low temperatures. 
Although we should be careful to check
outputs from such a complicated algorithm, 
the reported results, which are averaged over $100-400$ realizations
of disorder, is very attractive and likely to be decisive one
for this model. 
Replica Monte Carlo is also applied to higher dimensional
spin glass models~\cite{SW88}. The performance of cluster 
dynamics, however, seems not a remarkable one for these models.  
Houdayer~\cite{H01}
argued that the inability of cluster dynamics for 
three-dimensional spin glass models is a consequence
of a mismatch between the structure of phase space of the
model and the definition of a cluster in the algorithm.

\section{Designing Special Purpose Ensembles} \label{SS_new}

In the previous sections, we introduced
three algorithms, Exchange Monte Carlo,
Simulated Tempering, and Multicanonical Monte Carlo.
We also discuss cluster dynamics in Replica Monte Carlo.
In all cases, an ensemble with extension 
in the temperature or energy axis is useful 
in the sense that they have straightforward applications
to various problems in statistical physics. On the other
hand, we can introduce extensions specially designed for a target 
distribution and our computational aims. 
The use of such ensembles is 
already discussed in the pioneering
paper on Expanded Ensembles by Lyubartsev {\it et al.\/}~\cite{expand} and
in the studies on Adaptive Umbrella Sampling~\cite{M87,HVK92} (and
also in the earlier studies on Umbrella sampling, where the
weight is manually determined.). 
Here we discuss designing principle and utility of such 
``special purpose'' ensembles.\\

\noindent {\bf Complexity ladder}\\
To obtain a fast mixing Markov chain
and evaluate the free energy 
(multiple integrals), it is reasonable
to construct an ensemble composed of a sequence of distributions
that interpolates a ``complex/unknown'' distribution
to a ``simple/known'' distribution~\cite{expand,GT95,WL97,path}.
Wong~\cite{WL97} called such a structure as a
``{\bf complexity ladder}''. 
Here, the ``simple/known''
components should have sufficient entropy to obtain the
required diversity of paths to the ``complex/unknown'' states 
(The ferromagnetic
state is an example of the state
simple but does not have sufficient entropy.).
The canonical distributions with different temperatures
give an example of such structures, where
components of higher temperatures
correspond to simpler components. 
At the infinite temperature,
it reduces to a known distribution that described 
a completely disordered state. 
Another example is
provided by ``Multi-System-Size''-type ensembles
which consists of distributions of different system size $N$.
Here components with small $N$ fill the role of
simple components~\footnote{
It is clear that $N=0$ component has not sufficient entropy.
The intermediate states, however, can have enough entropy at moderate
temperatures. Algorithms with Multi-System-Size ensemble
will not be efficient or biased at very low temperatures where
these states have not enough entropy.}.
We will discuss them further in Sec.~\ref{SSS_msse}.
We can proceed more along this way. For example,
we can introduce an extended ensemble with ``soft spin''s.  
At an extreme, its component is a distribution
with discrete (or constrained) variables,
say, Ising spins or rigid plane rotators. At the other
extreme, its component reduces to a Gaussian distribution,     
whose samples are easily generated with
Cholesky decomposition of the covariance matrix.
This method is not implemented yet, but might be useful
for some hard problems in the study of spin glass and 
combinatorics.\\

\noindent {\bf Bridge}\\
A way to design artificial ensemble for efficient computation
is the inclusion of {\it non-physical configurations\/} (states
prohibited in the original problem)~\cite{expand,GT95}. 
The above-mentioned
soft-spin algorithm is regarded as an example.
Another example is ensembles for lattice polymers, which contain
conformations that violate the {\it self-avoiding condition\/}~\cite{expand,expand96,iba98,Chiken99}.
Specifically, Multi-Self-Overlap Ensemble (MSOE) introduced 
by Iba, Chikenji and Kikuchi~\cite{iba98} had a remarkable success 
in the calculation of ground states and thermodynamic properties
of lattice protein models~\cite{Chiken99}. Similar approaches
for off-lattice polymers with truncated Lennard-Jones cores are
discussed in~\cite{expand96,Chiken99,BD00,SB00}. 
The other examples are 
ensembles that relax {\it hard core condition}
of hard core fluid (solid)~\cite{expand,BWA,Mase99}~\footnote{
Hard core fluid is also treated by a multicanonical-type 
extension in the space of {\it volume} occupied by the fluid~\cite{SB96},
which is equivalent to an extension with 
{\it varying diameter of hard cores} in athermal models.
}
and ensembles for gene-propagation
analysis (pedigree analysis)~\cite{GT95} that contains the configurations
violating {\it Mendel's law} of genealogy.

These ensembles often result in great enhancement of 
the efficiency, because ``{\bf bridges}'' ~\footnote{
I borrow the key-word {\bf bridge} from Lin and Thompson~\cite{LT93}, while 
the term {\bf stepping-stones} have appeared in~\cite{JM96,Chiken99}.
{\bf Catalytic Tempering} algorithm 
proposed in~\cite{SB00} is also based on a
similar idea.}
provided by the non-physical configurations  
give a lot of additional 
paths between the configurations that are separated in the
original problem. 
An instructive example of such ``stepping stones'' is
shown in Fig.~1 of the reference~\cite{Chiken99}.
Of course, a drawback of such an approach is that we
``lose'' the non-physical samples. It is thus necessary
for the improvement in the mixing rate to be large enough
to overcome this loss.

For Multi-Self-Overlap ensemble for the
HP model of lattice heteropolymers, 
this requirement is checked by 
Iba~{\it et al.\/}~\cite{iba98}.
For a chain of length 56 with
highly degenerate ground states, simulation 
with Multi-Self-Overlap 
ensemble produces more independent samples
than a conventional Multicanonical Monte Carlo
within the same number of MCS, despite the 
loss of non-physical samples that violate 
the self-avoiding condition. 
Chikenji~{\it et. al\/}~\cite{Chiken99}
successfully applied Multi-Self-Overlap Monte Carlo   
to problems bio-physically more interesting,
{\it e.g.}, generation of the lowest energy state of 
a chain of length $100$ and calculation of 
thermodynamic properties of a protein-like chain of length $42$. 
Exploration of thermal states of such a long
heterogeneous chain have rarely been reported in this field.
Based on the results of these calculations,
Chikenji {\it et al.\/} proposed a hypothesis on the 
relation between the order of
the phase transition and ground state degeneracy.

Besides slow mixing of ``mathematically correct''
algorithms, genuine non-ergodicity of dynamics, i.e., 
the lack of the connectivity
of the graph defined by a transition matrix, is often
an annoying problem in Dynamical Monte Carlo.
It is not always easy to prove the ergodicity of 
a given Markov chain (We should carefully 
check unexpected appearance of isolated 
configurations. See  
Fig.~2.10-2.12 in~\cite{Sokal95} for examples of 
such configurations in self-avoiding walk simulations.).
The introduction of unphysical states as bridges 
provides a simple way to resolve the difficulty of non-ergodicity.
Examples are seen in the studies on self-avoiding 
lattice polymers~\cite{expand96,iba98,Chiken99},
pedigrees~\cite{GT95}, and  Latin squares~\cite{JM96}.

The inclusion of the forbidden states
as bridges is a natural idea to improve 
relaxation and has been used in
pre-extended-ensemble stages~\cite{ADD,LB93,JM96}. 
It is, however, not always easy to set
the penalty for 
putting adequate part of samples into ``bridges'' without
the idea of learning in preliminary runs (or the use of
multiple replicas). Now we can use any of three types of 
implementations for this purpose, i.e.,
Exchange Monte Carlo, Simulated Tempering, and
Multicanonical Monte Carlo.
The extended ensembles with non-physical
bridges are also considered as finite temperature version
of constrained optimization algorithms 
by Geman {\it et al.\/}~\cite{Geman90}. \\

\noindent {\bf Calculation of a free energy surface}\\
An important motivation to special purpose
ensembles is the observation 
of rare events and calculation
of free energy surfaces.
When we want to calculate free-energy surface 
as a function of one- or two- macroscopic variables,
we can use an ensemble extended in the dimension
of these variables~\cite{M87,HVK92,BHN93,BK97,HIGO97,SLV97}. 
For example, Sheto 
{\it et al.\/}~\cite{SLV97} design an ensemble
for the study of the free-energy surface 
of an Ising model on the 
energy-magnetization plane.
Similar approaches are extensively used in the studies with
Adaptive Umbrella Sampling~\cite{M87,HVK92,BK97}
for the calculation
of free energy as a function of 
reaction coordinates ({\it e.g.}, dihedral angles of polypeptides).
{\bf Multioverlap ensemble}~\cite{MOE} designed for the calculation of
the distribution of ``replica overlap'' between two independent
samples is also based on a modification of this strategy.

Note that the extension required for the measurement often does not
provide large entropy states that are necessary to
fast mixing of the Markov chain. 
For this reason, some of the 
calculations by Adaptive Umbrella Sampling
or Multioverlap ensemble might possibly be affected by the
slow mixing of the Markov chain.  It is often 
covered by fast tunneling between the states that
have extreme values of the reaction coordinate, but
such tunneling does not ensure unbiased sampling
from all metastable states.
Multi-dimensional extensions
provide a way to circumvent this difficulty.
We will discuss them in the next paragraph.  \\

\noindent {\bf Multivariate extensions}\\
To implement special purpose ensembles, the idea of
ensembles extended in multi-dimensions (multivariate/multi-component
extensions, \cite{SLV97,BK97,HIGO97,NN98,iba98,
Chiken99,CK00,Hatano,SKO00}) 
plays an important role.
Since the use of too many dimensions is not realistic,
bivariate or three-variate extensions are usually most useful.
For example, in the case of Multi-Self-Overlap ensemble for 
lattice polymers~\cite{iba98,Chiken99}, we use 
an ensemble defined by uniform density 
on two-dimensional space {\sf (degree of self-overlap, energy)}.
It seems essential for heteropolymers with attractive interactions
to include the extension in the space of the energy, 
because the relaxation of the self-avoiding constraint   
often causes collapse of polymers at low temperature. 

As we have already remarked in the previous paragraph,
two-dimensional extension is especially useful for
extended ensembles for the measurement
of rare events (see Sec.~\ref{SS_art}). That is,
two-dimensional extensions, say, {\sf (an axis 
for the measurement, the axis of the
temperature)}
or {\sf (an axis for the measurement, the axis of 
the energy)} improve the efficiency (and safety) of 
the algorithms for the calculation 
of free energy surfaces on the axis of measurement. 
Examples are found in~\cite{SLV97,Hatano,HIGO97,CK00,SKO00}. 

For example,
Chikenji and Kikuchi~\cite{CK00}
explored the entropy density of a lattice
protein model (a G\={o} model) using an extension of 
Multi-Self-Overlap 
ensemble defined by the uniform density 
in a three-dimensional space spanned by 
{\sf (the degree of self-overlap, the number
of native contacts, the number of total contacts)}.
Their motivation is the study of the curious 
folding mechanism of $\beta$-lactoglobulin, where 
$\alpha$-helices rich intermediates tentatively appears
before it finally folds into a stable $\beta$-sheet rich 
conformation. Entropy density on the space
{\sf (the number
of native contacts, the number of total contacts)}
calculated by the method vividly
illustrates the role of entropy in the folding.

How can we construct ensemble with multi-dimensional extensions?
We have already discussed it for the case of multicanonical
ensembles (Sec.~\ref{SS_multi}). For Exchange Monte Carlo and
Simulated Tempering, the introduction of a multi-parameter
family is also straightforward. 
In the case of Exchange Monte Carlo~\cite{iba98,H99,SKO00},
a two-dimensional version of simultaneous distributions 
of replicas is expressed as
\begin{equation}
\label{S2dim}
\tilde{p}(\{\boldx_{kj}\})= \prod_{(k,j) \in G}  p_{kj}(\boldx_{kj}) {}_.
\end{equation} 
where the values of two parameters are indexed by $k$ and $j$. 
There is a degree of freedom in the
choice of a Graph $G$ that defines
the way of the extension. A way is the use of a ``lattice type''
configuration of $(k,j)$ in the two-dimensional parameter
space~\cite{SKO00}. Another possibility 
is the use of a ``quasi one-dimensional''
configuration where $(k,j)$ are points on a curve in the
parameter space~\cite{H99}. The former corresponds 
to two-dimensional multicanonical
algorithms. The latter saves computational resources, 
but it is not easy to tune the large degree of freedom 
of setting a curve in the two-dimensional space.

\section{Multi-System-Size Ensembles}
\label{SSS_msse}

As we have discussed in the previous sections,
we can freely choose the space of extension 
in Extended Ensemble Monte Carlo. An interesting 
possibility is the use of an extension 
in a ``space of size of the system'' (or, in general,
in a ``space of dimensionality of the state space''),
which corresponds to 
sampling from a mixture of systems of different sizes.
A simulation that realizes such an ensemble is considered
as a ``growth/diminish (construction/destruction) method''
for sampling a probability distribution. 
Note that monotonic growth of the system 
is prohibited by the detailed-balance condition.

In the paper~\cite{msse}, I discuss a simulation of 
spin glass with an ensemble extended in the space of
system-size, Multi-System-Size Ensemble. After I completed
the work, I came to know references with similar ideas
in various fields of physics and statistical sciences.
Here, we will give
a {\it cross-disciplinary\/} survey on this subject.

We begin with simulations of fluids with a varying number
of particles. We know an ensemble
corresponding to such a simulation. It is 
{\bf grandcanonical ensemble} found in any textbook
of statistical physics. It is interesting to 
introduce multicanonical or other types of 
extensions in the space of particle number 
(particle density, chemical potential)
as a generalization of 
grandcanonical ensemble.  
While the number of particles
fluctuates in a limited range 
in conventional grandcanonical ensemble,
it can fluctuate in an arbitrary wide range, say, zero to 1000
in the extended ensemble.
In the literature, Lyubartsev {\it et al.\/}~\cite{expand} 
already discussed it in the context of free energy 
calculation. An application to Lennard-Jones fluid is found in 
Wilding~\cite{Wnum}, which explored the
subcritical coexistence region of Lennard-Jones fluid 
by a multicanonical ensemble defined by the uniform
density on the space of the particle density.
 
There is a variation on this idea.
When we are interested in the calculation of
the chemical potential, 
an ensemble which gives an interpolation
between size $N$ and size $N+1$ systems is often required. 
It is precisely realized by an extended ensemble that
contain components corresponding to systems with {\it partial
decoupling\/} between a particle and other $N$ particles.
Examples of such a ``ghost particle''(ghost polymer) method
are found in~\cite{WM94}
and~\cite{LFL98}. 
In the latter study~\cite{LFL98}, 
solvation free energies of methane and 
alkali halide ions are calculated, and 
the results are compared to experimental data.
These ensembles are also considered as examples of 
the extended ensembles that consist of non-physical systems. 
Similar idea with an umbrella method is also 
found in~\cite{DV93}.

Another application in physics where extension
in the space of system-size is naturally introduced
is simulation of the polymers. 
It is considered
as a variation of grandcanonical simulation of 
self-avoiding walk~\cite{Sokal95}, where monomers
are added and removed at the both ends of a 
polymer chain. In the conventional grandcanonical simulation,
the length of polymers fluctuates around an equilibrium length.
Introducing the idea of extended ensembles, we can systematically 
enhance the fluctuation and perform practical calculations
for intricate models, such as random walks in a 
restricted geometry and with complex interactions 
between monomers.   
For example,
an adaptive, multicanonical-like method for self-avoiding
walk is given by Grassberger~\cite{G93}, while 
grandcanonical simulation with a length dependent 
chemical potential is already seen in earlier works, 
{\it e.g.,}~\cite{RW91}.  In \cite{G93}, self-avoiding
walks on a lattice with random obstacles (i.e., randomly chosen
excluded sites) up to the length $100$ are efficiently generated
by the method, and it is shown that the universality class
of the problem is different from the one of the corresponding 
uniform lattice. Escobedo and Pablo~\cite{EP95,EP96}
reported a systematic study of 
expanded ensemble simulation of polymers where the
length of polymers are dynamically changed. 
Their method is recently applied 
to diblock copolymer~\cite{WYNP00}.
It might be interesting to point out that 
a preliminary work on multicanonical ensemble~\cite{B87}
also treated a size-variable system, an 
ensemble of random surfaces.

Extended Ensemble Monte Carlo algorithms 
with a state space of
varying dimension is also of current interest
in statistical sciences. 
It is natural to apply them to the problems
with a built-in sequential structure, say, time-series
modeling and gene-propagation analysis. 
On the other hand, they
have an interesting motivation in statistics,
i.e., they are useful for the simulation of 
a (posterior) distribution over the space of models
with different number of parameters~\cite{jump,Doucet01}
(Here a ``parameter'' 
means an element of a model to be estimated
from data.)~\footnote{
We note that models  
with a large number of parameters do not necessarily 
show better performance with finite number of data.
It is rather evident when we
consider extreme cases, such as fitting of 10
data points by 9th order polynomials and
classification of 100 samples into 100 clusters.
Thus we need to select a model (or mix models)
with an appreciate number of parameters.}. 
After a pioneering comment by
Wong~\cite{Wong95},
Wong and Liang~\cite{WL97} gives
an application of their idea in 
various types of the problems, 
including a Traveling Salesman Problem.
Liu and Sabbati~\cite{LS98} extensively discussed 
applications of ``{\bf Simulated Sintering}''
method in statistics~\footnote{In these works, 
a ``Dynamic Weighting'' technique
proposed in~\cite{WL97,LW99} is used, which does not belong
to the class of Extended Ensemble 
Monte Carlo defined in this paper.}.

Finally, we will touch on a related idea, extended ensemble
that consists of systems of {\it variable types 
of dynamical variables}.
Kerler and Weber~\cite{KW93} discussed
an extended ensemble simulation of Potts model, where 
the number $q$ of possible states of a spin (``colors'') 
is dynamically changed in a run. They implement the idea
with a combination of cluster dynamics. 
A corresponding situation in statistics is 
found in Richardson and Green~\cite{RG97},
which deals with simulation-based Bayesian classification
of objects to an unknown number of clusters~\footnote{The work
does not seem to use an iterative learning procedure to improve
the sampling scheme. In this sense, it is not Extended Ensemble
Monte Carlo defined in this paper.
A reason that we refer to the study here is
that it provides a good example of the application of Dynamical Monte Carlo 
in computational statistics. 
}.
They treated problems like ``How many Gaussian components are
identified in a given experimental data ?'', and
designed a Monte Carlo procedure for computation of probabilities.
In their work, the cluster
to which each object belongs is indicated by the state of 
a Potts spin corresponding to the object. 
Then, dynamical changes of the number $q$ of the clusters
corresponds to variation of the 
number $q$ of the states of Potts spins.

\section{Summary and Future Perspectives}

In this paper, we review three types of
Extended Ensemble Monte Carlo algorithm,
i.e., Exchange Monte Carlo, Simulated Tempering,
and Multicanonical Monte Carlo and discuss approaches
with special purpose ensembles. We also give a guide to 
extended ensembles with variable dimension of state spaces 
and Replica Monte Carlo algorithm.  Throughout the paper, the
possibility of various type of extensions is
stressed.  They are not only useful for the calculation
of free energy, but also efficient for acceleration of
the relaxation. Our idea is summarized as the
following ``correspondence principle'':
\begin{quote}
{\it
If we have an annealing strategy for searching 
ground states, 
we can design an Extended Ensemble Monte Carlo algorithm
to sample from the corresponding distribution.
}
\end{quote}
With this principle, we can translate optimization
algorithms to algorithms for the calculation of thermodynamic
properties. Note that this is only true for
simulated annealing-type
algorithms, and not applicable to more
intricate/sophisticated optimization algorithms, {\it e.g.}, 
algorithms with ``genetic crossover'' or with other heuristics
that violate detailed-balance, methods based on the 
use of the correspondence between ground states of different systems. 
The principle, however, still provides a useful
guideline for the construction of sampling schemes.

With Extended Ensemble Monte Carlo algorithms,
we can attack difficult problems where conventional
Dynamical Monte Carlo algorithms are too slow
even with the most powerful computers.
Up to now, the most significant applications are
found in computational physics and statistical information
processing. But I believe that Extended Ensemble Monte Carlo is 
also a key to any field that requires sampling from
complex distributions and estimation of the entropy.
The introduction of Dynamical Monte Carlo -- 
in 1950s for physics and in 1990s for statistics
-- gave great impacts on these fields.
I hope that Extended Ensemble Monte Carlo
will give {\it second impacts\/} on 
the study of the fields
where we are interested in the properties of probabilistic
distributions and {\it large deviation\/} from non-weighted averages,
including computational physics and statistics
as a special example.

\newpage

\section*{Acknowledgement}

I would be grateful to Dr.~K.~Hukushima, Prof.~M.~Kikuchi, Mr.~G.~Chikenji,
Prof.~Y.~Okamoto, Prof.~H.~Nakamura and
Dr.~A.~Doucet for giving me kind advices and teaching references.

\newpage

\appendix

\section*{Comparison of the Methods} \label{SS_appendix}

In this appendix, we discuss the issues on relative advantages
of the three (types of) algorithms,  
Exchange Monte Carlo 
(Metropolis-Coupled Chain, Parallel Tempering), 
Simulated Tempering (Expanded Ensemble Monte Carlo), and
Multicanonical Monte Carlo (Adaptive Umbrella). 
The results are summarized in the
following table: 

\begin{flushleft}
\begin{tabular}{|l|l|c|c|c|} \hline
 Subjects & Com- & Exchange & Simulated  & Multicanonical \\ 
 &  ment &    Monte Carlo & Tempering  &  Monte Carlo \\ 
 & \# &  [Sec.\ref{SS_exg}] & [Sec.\ref{SS_temp}] & [Sec.\ref{SS_multi}] \\ 
\hline \hline
First Order Trans. & 1 &
$\times$ & $\times$ & $\bigcirc$  \\ \hline
Non Exp. Family & 2 &
$\bigcirc$ & $\bigcirc$ & $\times$ \\ \hline
Replica Overlap & 3 &
$\bigcirc$ & $\times$ & $\times$ \\ \hline
Learning Speed & 4 & 
$\bigcirc$ & ? & ?  \\ \hline
Molecular Dynamics & 5 & 
$\bigcirc$ & $\bigcirc$ & $\bigcirc$  \\ \hline
Step Size Control & 6 &
$\bigcirc$ & $\bigcirc$ & ? \\ \hline
\end{tabular}
\end{flushleft}

\vspace{0.3cm}
\noindent
While the symbol $\bigcirc$ indicate that 
the algorithm has an advantage on the subject,
$\times$ means that the algorithm has severe disadvantage on the
subject. 
The symbols $?$ means ``still controversial''.
The number in ``comment \#'' column indicates 
the item number~(\#) of the discussion 
on the corresponding subject. 

\vspace{0.3cm}

\noindent
We will give remarks on the subjects in
the table in the following:

\begin{enumerate}
\item {\bf First Order Transition} 

As we have discussed in the previous sections, a remarkable advantage
of Multicanonical Monte Carlo is that it can treat 
systems with first-order-like
transitions. 

\item {\bf Non-Exponential Family}

On the other hand, non-exponential 
family of distributions is 
not suitable for multicanonical-type treatment.
While the significance of non-exponential 
family is not clear in statistical physics,
they are often 
important in the applications in statistical sciences.     
\newpage

\item {\bf Calculation of Replica Overlap}

In the study of statistical physics of random systems,
it is often required to calculate the distribution of 
a quantity defined with two independent 
samples from a Gibbs distribution.
An example of such quantities is 
``replica overlap'' $q$, which is defined as
an overlap of mutually independent samples 
$\boldx$ and $\boldx^\prime$ from a given 
distribution. An easy way to compute 
such a quantity is to simulate two 
independent Markov chains $S$, $S^\prime$
and use a pair $(\boldx, \boldx^\prime)$ of samples
where $\boldx$ and $\boldx^\prime$ are sampled 
from $S$ and $S^\prime$ respectively.
Then, independence of
$\boldx$ from $\boldx^\prime$ is assured even
with slow mixing of the Markov chains $S$ and $S^\prime$.

This simple method, however, does not work well 
when we use Simulated Tempering, because the states in 
two chains have different
values of temperature for most part of the simulation 
--  they coincide with each other with probability $1/K$
when the number of discretized temperatures
is $K$, which results in severe waste of samples. 
The situation is essentially the same when we use multicanonical-type
algorithms, or, when any parameter is used for the 
construction of the extended ensemble. 
For Simulated Tempering, we can use two copies of the system 
with a common temperature variable, but it lowers the performance
the algorithm. A few other 
methods have been proposed up to now, 
but all of them have some drawbacks, {\it e.g.},
they cause slow relaxation of system (Berg and 
Celik~\cite{BC92b})
or introduce additional complexity (multioverlap ensemble~\cite{MOE}).

On the other hand, the calculations of 
the overlap $q$ and other
quantities defined with two independent 
samples from the original distributions are straightforward
with Exchange Monte Carlo algorithm.
We just simulate two independent Markov chains 
each of which consists of 
$K$ replicas with the same set of temperature $\{\beta_k\}$.
Then we calculate and record the overlap of 
two replicas with the same temperature whose
time evolution is governed by mutually independent 
Markov chains.
This is a significant advantage
of Exchange Monte Carlo algorithm.

\item {\bf Learning Speed in Preliminary Runs} 

Exchange Monte Carlo algorithm seems to show better performance
and less complexity in the learning phase. It does not
require simultaneous tuning of the strength of the penalty  
and discretization of the temperature required in Simulated Tempering.
The tuning of the values of 
the temperature (parameter) of replicas is still required, but
we can enjoy a benefit from the use of the exchange rate
between replicas. 

On the other hand,  
Simulated Tempering and Multicanonical Monte Carlo
have a handicap in the learning stage, if we use the naive method
of tuning based on the frequency of the visits to a temperature 
or an energy.
It is because a random walk on the temperature or energy axis
causes fluctuation of the visiting frequencies, which 
directly induces instability of the calculated weights.

Some authors~\cite{Smith,SB96,KR94}, 
however, proposed tuning methods based on the
acceptance ratio or the ratio of frequencies, 
which will reduce the instability of this type. Recent
development of Flat Histogram Monte Carlo and related
algorithms~\cite{Oliv,Wang00,Wang01} can also improve 
the efficiency of the learning stage. 
On the other hand, experiences on difficult
cases suggest that the most difficult part of the tuning
phase is often the determination of several weights near ground states
of the system (and ground states themselves).
We need more experiences and carefully designed
experiments to evaluate these factors -- This is
the reason why we give ``?'' to this subject.   

\item {\bf Molecular Dynamics, Hybrid Monte Carlo, Langevin Equation}

{\bf Molecular Dynamics} is a useful tool
for the simulation of continuous systems, say,
simulation of realistic protein models, 
even when we are interested only in equilibrium
properties. Specifically, combinations of Dynamical Monte Carlo 
and Molecular Dynamics ({\bf Hybrid Monte Carlo}) are 
convenient tools for the sampling from Gibbs distributions.
There are also methods based
on Langevin Equation, which can be regarded
as a version of Hybrid Monte Carlo.

Here we consider how to combine these methods with the idea of 
Extended Ensemble Monte Carlo.
At first sight, Exchange Monte Carlo and Simulated Tempering
have an advantage, because implementation
is quite straightforward~\cite{LFL98,expand96,IrbackMD,
latexchange,SO99,CNFD00}. 
For example, 
the addition of an exchange procedure is 
enough for the combination of
Exchange Monte Carlo and Hybrid Monte Carlo, where
the states of replicas are swapped with fixed values of the 
corresponding demons (i.e., momentum part of the Hamiltonian)~\footnote{
Sugita and Okamoto~\cite{SO99} have proposed a different method, where
demons and replicas are exchanged simultaneously 
with rescaling of momenta of the demons. Another approach is
introducing an exchange procedure to microcanonical ensembles in which 
demons are integrated out~\cite{CNFD00}.}.

On the other hand, multicanonical-type implementation requires
the estimation of the derivative $d \log D(E)/d E$ of the logarithm of the 
density of state, which causes additional
complexity in the tuning stage. However, studies by 
several authors~\cite{HVK92,Hans96,NN97,NN98,BK97,BK98,lathybrid} have proved 
that it is not difficult despite
the apparent difficulty. 

\item {\bf Step-Size Control}

For continuous systems, the step-size of trial moves (or, in general,
the distribution of the sizes and directions of trial moves) 
is an important factor in the mixing of the Markov chain
governed by Metropolis dynamics.
The optimal step-size depends on the temperature
and other parameters of the target distributions.
There is no established way for the determination 
of optimal step-size, but there are some ``rules of thumb'',
{\it e.g.}, step-size with moderate 
acceptance ratio (say, $\sim$ 50\verb+%+)
usually gives good results.

For Exchange Monte Carlo and Simulated Tempering,
step-size can depend upon the temperature (or
any parameter used for the construction of extended ensemble.).
It does not spoil the detailed balance condition
because the flips in a replica (the flips with a fixed temperature) 
are separated from replica exchanges (temperature changes)
in these algorithms. 

It is not the case with
Multicanonical Monte Carlo.
If we use energy dependent step-size
in a multicanonical simulation of a continuous system,
it usually gives wrong results, because it 
results in violation of the detailed balance condition.
The tuning of the weights that compensate the
bias caused by energy dependent step-size, if possible,
seems complex and not realistic.

This disadvantage, however, can be solved by using
a ``patchwork'' of ensembles proposed by several
authors~\footnote{
Hansmann~\cite{Hans97} discussed a 
patchwork of Tsallis ensembles. 
A patchwork of locally multicanonical ensembles
is discussed by Sugita and Okamoto~\cite{SO00,Mitsutake01}.
The present author (Y.~Iba) have
also proposed a version of Exchange Monte Carlo  
algorithm, where canonical ensemble is replaced by
the ensemble defined by $E<E_{th}$ (unpublished).
Similar ensembles naturally appear, when
demons (momentum part of the Hamiltonian) 
are integrated out from 
microcanonical ensembles~\cite{CNFD00}.
}. 
For example, consider an ensemble composed by several
multicanonical-type ensembles, each of which is 
defined on an interval of the energy axis. These intervals
are assumed to partly overlap each other and we use
a method like Exchange Monte Carlo for the
integration of them.
In this setting, we can safely use different step sizes
in different components.

\end{enumerate}

\end{document}